\documentclass[aps,prl,a4paper,twocolumn,superscriptaddress,10pt,showpacs]{revtex4-1}

\usepackage[pdftex]{hyperref}
\usepackage[pdftex]{graphicx}
\usepackage{amsmath,amsfonts,amssymb,bm}

\newcommand{\be}{\begin{equation}}
\newcommand{\ee}{\end{equation}}

\newcommand{\aIR}{\alpha} 
\newcommand{\avdn}[1]{\avg{\delta n}{}^{(#1)}}
\DeclareMathOperator{\PV}{P}	% Cauchy principle value

\newcommand{\vc}[1]{\bm{#1}}
\renewcommand{\r}{{\vc r}}
\renewcommand{\k}{{\vc k}}
\newcommand{\q}{{\vc q}}
\newcommand{\p}{{\vc p}}

\newcommand{\V}{\mathcal{V}}

\newcommand{\lloc}{l_\text{loc}}

\newcommand{\rmd}{\mathrm{d}}
\newcommand{\intdd}[1]{\int \frac{\rmd^d #1}{(2\pi)^d}}	% k-space integral
\newcommand{\intddr}{\int \rmd^d r}			% realspace integral

\newcommand{\gh}[1]{\hat\gamma_{#1}}		% quantized Bg amplitude
\newcommand{\ghd}[1]{\hat\gamma^\dagger_{#1}}	%

\newcommand{\epn}[1]{\epsilon^0_{#1}}		% free kinetic energy
\newcommand{\ep}[1]{\epsilon_{#1}}		% Bogoliubov energy

\newcommand{\avg}[1]{\overline{{#1}}}	

\newcommand{\xpct}[1]{\bigl\langle #1 \bigr\rangle} 
\newenvironment{smallpmatrix}{\left(\begin{smallmatrix}}%
{\end{smallmatrix}\right)}
\newcommand{\matrs}[4]{\begin{smallpmatrix}
#1 & #2 \\
#3 & #4 
\end{smallpmatrix}
}

\begin{document}

\title{Bogoliubov Theory of Disordered Bose-Einstein Condensates}
\author{Christopher Gaul}
\affiliation{Departamento de F\'isica de Materiales, Universidad Complutense, E-28040 Madrid, Spain}
\author{Cord A.\ M\"uller}
\affiliation{Centre for Quantum Technologies, National University of Singapore, Singapore 117543, Singapore}
%\date{\today} 

\begin{abstract}
We describe interacting bosons at low temperature %Bose-Einstein condensates 
in spatially correlated random potentials. 
By a Bogoliubov expansion around the deformed mean-field condensate,
the fundamental Hamiltonian for elementary
excitations is derived, achieving an analytical formulation in the case of weak disorder. 
From this, we calculate the 
sound velocity and true zero-temperature condensate depletion 
in correlated disorder and all dimensions.
\end{abstract}

\maketitle

The interplay between interaction, quantum statistics, and
randomness defines one of the richest problems in 
condensed matter physics: the dirty boson problem \cite{Giamarchi1988,Fisher1989}.
Determining the ground state of a 
disordered Bose gas is already a formidable task 
\cite{Pollet2009}; 
all the more desirable is an analytical theory for the excitations of disordered Bose-Einstein
condensates (BECs). 
In presence of a well-developed condensate, i.e.\ at low temperature
and for weak disorder, the most economic description, due to
Bogoliubov \cite{Bogoliubov1947}, treats quantum fluctuations around
the best mean-field approximation to the condensate. 
One of the key quantities that such a theory should provide is
the experimentally measurable, 
\emph{disorder-renormalized sound velocity}, which characterizes the
low-energy dispersion and enters all thermodynamic
properties. 
In this respect, existing theories for bosons in disorder
are not entirely satisfactory.
While simple approaches cannot determine a change in 
quasiparticle dispersion at all 
\cite{Huang1992,Kobayashi2002,Astrakharchik2002}, more elaborate
calculations find a positive correction to the sound velocity
due to uncorrelated disorder \cite{Giorgini1994,Lopatin2002,Falco2007}.
Still others, in different settings and with different methods, 
report a negative correction
\cite{Lee1990,Zhang1993,Yukalov2007,Gaul2009a}. 
Here, a unifying framework is missing.
On a more conceptual level, the  
\emph{condensate deformation} 
by disorder is often confounded with     
the  \emph{condensate depletion}, i.e.\ the fraction of particles not
in the condensate at all. 
This quantum depletion is a 
crucial quantity whose smallness validates Bogoliubov's approach; it contains all particles with non-zero
momentum \cite{Pitaevskii2003} \emph{only if} the 
condensate is homogeneous. 
If the condensate is deformed, it has
non-zero momentum components on its own, already 
on the mean-field level
\cite{Sanchez-Palencia2006}.  
Only these have been counted by previous approaches
\cite{Huang1992,Kobayashi2002,Astrakharchik2002,Giorgini1994,Lopatin2002,Falco2007}.
For the true disorder-induced quantum
depletion there are, to our knowledge, only few numerical
results \cite{Singh1994}. An analytical calculation of this pivotal quantity 
is lacking so far. 

With this Letter, we present a comprehensive Bogoliubov theory for inhomogeneous
Bose-Einstein condensates that 
gives access to a wealth of relevant
quantities, including the full excitation dispersion with the
renormalized speed of sound and the localization length of elementary excitations
\cite{Bilas2006,Lugan2007a}. 
Notably,
our formulation encompasses the class of spatially correlated
disorder that is currently under study with ultracold gases 
\cite{Sanchez-Palencia2010,Modugno2010,Dries2010}. 
By expanding the
many-body Hamiltonian around the deformed mean-field solution, 
we derive the fundamental Bogoliubov Hamiltonian for 
excitations, verified to be orthogonal to the ground state.  
For weak disorder, a fully analytical description is
achieved. From this Hamiltonian, we calculate 
corrections to the sound velocity and zero-temperature quantum
depletion for correlated disorder in all dimensions.
 
%----------------------------------------
\section{Starting line} 

Interacting bosons are described by 
\begin{align}\label{eqManyParticleHamiltonian}
 \hat E = \int &{\rm d}^d r\,
\hat \Psi^\dagger(\r) \left[h(\r) + \frac{g}{2}\hat \Psi(\r)^\dagger
 \hat\Psi(\r)
 \right] \hat \Psi(\r),
\end{align}
with the grand-canonical single-particle Hamiltonian  
\begin{equation} 
h(\r) =-\frac{\hbar^2}{2m}\nabla^2+V(\r)-\mu
\end{equation} 
 and  field operators that obey
$\bigl[\hat \Psi(\r) , \hat \Psi^\dagger(\r')\bigr] =
\delta(\r-\r')$
\cite{Pitaevskii2003}.
The global confining potential is assumed to be very smooth, 
ideally a very large box, and  $V(\r)$ describes local spatial fluctuations.
Repulsive interaction between bosons is accounted for by $g = 4\pi
\hbar^2 a_{\rm s}/m >0 $ (in $d=3$, with s-wave scattering length $a_{\rm s}$), an excellent approximation for 
cold and dilute gases, where the gas parameter $(n a_{\rm
  s}^3)^{1/2}$ is small. 

Below a critical temperature, the Bose gas forms a BEC, 
i.e.\ a large fraction of
particles condense into the ground state of the single-particle density
matrix. In the absence of interaction, this is the ground state
of the potential $V(\r)$. Also interacting bosons condense, into
a mode  whose shape results from 
the competition between kinetic energy, confinement and interaction. 
Bogoliubov theory \cite{Bogoliubov1947} takes advantage of this
macroscopic  occupation and splits the quantum field into a mean-field condensate and quantized fluctuations: 
$\hat \Psi(\r) = \Phi(\r) + \delta\hat\Psi(\r).$

%----------------------------------------
\section{Condensate}

We first describe how a weak external
potential deforms the condensate. By definition, the
ground state 
 minimizes the energy \eqref{eqManyParticleHamiltonian} on the
 mean-field level and thus obeys the stationary 
Gross-Pitaevskii (GP) equation
%\begin{equation}\label{GPeqn}  
$h(\r)\Phi(\r) + g \Phi(\r)^3 =0$  \cite{Pitaevskii2003}. 
%\end{equation} 
The condensate's kinetic energy is 
minimized by choosing a fixed global phase, and we can take
$\Phi(\r)$ real. 
The imprint of a weak potential on the condensate amplitude can be computed 
perturbatively by  expanding  
$ \Phi(\r)=  \sqrt{n} + \Phi^{(1)}(\r)+ \Phi^{(2)}(\r) + \ldots$ in powers of $V$
around the homogeneous solution  $\Phi^{(0)} = \sqrt{n}$ \cite{Lee1990,Sanchez-Palencia2006}.
In order to maintain a fixed average density $n$, 
also the chemical potential is adjusted at each order, 
$\mu = gn + \mu^{(1)} + \mu^{(2)} + \dots$. 
We insert these expansions into the GP equation, transform
to momentum representation, and
collect orders up to $V^2$. 
The first-order imprint then is 
\begin{equation}\label{eqSmoothingPsi1}
\Phi^{(1)}_\q  = -\frac{ (1-\delta_{\q 0})V_\q}{ 2 g n +\epn{q}} N^{1/2} . 
\end{equation} 
This linear-response deformation is proportional to the potential's matrix element
$V_\q = L^{-d} \intddr
e^{-i\q\cdot\r}V(\r)$.
%; \new{the factor $(1-\delta_{\q0})$ guarantees
%  that the average density be fixed at $n$ to first order in $V$ even for a potential with non-zero mean.} 
In the denominator of eq.~\eqref{eqSmoothingPsi1}, the bare kinetic
energy $\epn{q}=\hbar^2q^2/2m$ equals the interaction energy $gn$ when
$q^{-1}$ equals the BEC healing length $\xi = \hbar/\sqrt{2 m g
  n}$. Thus, the condensate readily follows potential components with
$q\xi\ll1$, but shows a strongly smoothed imprint when  
$q\xi\gg1$ \cite{Sanchez-Palencia2006}. 
Pushing the expansion to
second order, we find 
\begin{equation}\label{eqSmoothingPsi2}
 \Phi^{(2)}_\q
   = \frac{1}{N^{1/2}} \sum_\p \Phi^{(1)}_{\q-\p} \Phi^{(1)}_{\p}  \frac{ (1-\delta_{\q0})\epn{p} -g n}{2 g n + \epn{q}} .
\end{equation}
Eqs.~\eqref{eqSmoothingPsi1} and \eqref{eqSmoothingPsi2} determine the
disorder imprints also on derived quantities like the
density $n_\k = L^{-d} \sum_\q \Phi_{\k-\q}\Phi_\q$.

%----------------------------------------
\section{Fluctuations} 

We expand the Hamiltonian \eqref{eqManyParticleHamiltonian} in powers of $\delta
\hat\Psi$ and $\delta \hat\Psi^\dagger$  around the condensate.  
The linear term vanishes, because $\Phi(\r)$ minimizes the energy functional. 
The relevant contribution is then the quadratic part that can be 
readily expressed in density-phase variables
$\delta\hat\Psi =  \delta \hat n(\r)/2 \Phi(\r) 
         +  i \Phi(\r) \delta\hat\varphi(\r)$ 
\cite{Gaul2008}: 
\begin{align}\label{eqBogoliubovHamiltonian}
\hat H  
= \int {\rm d}^d r \biggl\lbrace 
\frac{\hbar^2}{2m} \biggl[  &\left(\boldsymbol\nabla \frac{\delta \hat n}{2\Phi(\r)} \right)^2
+ \frac{\left[\nabla^2 \Phi(\r)\right]}{4 \Phi^{3}(\r)} \delta \hat n^2 \nonumber \\
& + \Phi^2(\r) (\boldsymbol{\nabla} \delta \hat \varphi)^2\biggr] + \frac{g}{2} \,\delta \hat n^2 \biggr\rbrace .
\end{align}

In a homogeneous system
with $\Phi(\r)=\sqrt{n}$, it is advisable to transform to Fourier
space and Bogoliubov excitations   
\begin{equation} \label{gammakdef} 
\gh{\k} = 
 \frac{\delta \hat n_\k}{2a_k\sqrt{n}} + i a_k \sqrt{n} \,\delta \hat \varphi_{\k}. 
\end{equation}  
These bosonic excitations obey 
$[\gh{\k},\ghd{\k'}]=\delta_{\k\k'}$. 
Moreover,  by choosing $a_k =
(\epn{k}/\ep{k})^{1/2}$ 
the homogeneous Hamiltonian becomes diagonal, 
\begin{equation} 
\hat H^{(0)} = \sum_{\k} \, \ep{k} \ghd{\k}\gh{\k},
\end{equation} 
with $\ep{k} =[\epn{k}(2gn+\epn{k})]^{1/2}$ the clean Bogoliubov
dispersion relation \cite{Bogoliubov1947}. Characteristically, low-energy excitations ($k\xi\ll1$) have the  
linear dispersion  
$\ep{k} =\hbar c k$, with sound velocity
$c=\sqrt{gn/m}$.

In presence of disorder, two things change. First of all, the reference point, from
where the excitations originate, is shifted  
to the inhomogeneous ground state $\Phi(\r)$. But still,  
eq.~\eqref{gammakdef} is fit to define excitations with the proper commutation
relations. The density-phase representation \eqref{gammakdef} implies
also that the fluctuation  
\begin{equation}  \label{eqBasedOnHydro} 
\delta\hat\Psi(\r) =  \sum_{\k} [u_\k(\r) \gh{\k} -
  v_{\k}(\r)^*\ghd{\k}] 
\end{equation}
decomposes over excitations with modes 
\begin{align} 
u_\k(\r) & = \frac{1}{2L^{d/2}}\left[\frac{\Phi(\r)}{a_k\sqrt{n}}+ \frac{a_k\sqrt{n}}{\Phi(\r)} \right]
e^{i\k\cdot\r} \label{modeu},  \\ 
v_\k(\r) & = \frac{1}{2L^{d/2}}\left[ \frac{\Phi(\r)}{a_k\sqrt{n}}- \frac{a_k\sqrt{n}}{\Phi(\r)}\right]
e^{i\k\cdot\r}. \label{modev}
\end{align} 
For the clean system with $\Phi(\r) = \sqrt{n}$, these modes reduce to
plane waves with the well-known amplitudes $u_k = \frac{1}{2}(a_k^{-1}
+ a_k)$ and  $v_k = \frac{1}{2}(a_k^{-1}
- a_k)$ such that $u_k^2- v_k^2=1$. In presence of disorder, the modes
\eqref{modeu} and \eqref{modev} are defined such that they still satisfy 
the bi-ortho\-gonality 
\begin{equation} 
\int {\rm d}^d r \left[ u^*_\k(\r) u_{\k'}(\r) - v^*_\k(\r)
  v_{\k'}(\r) \right] = \delta_{\k\k'}
\end{equation} 
required for
eigenmodes of the Bogoliubov Hamiltonian.  Moreover, the excitations are also orthogonal to the deformed
ground state, because $\Phi(\r) \left[u_\k(\r) - v_\k(\r) \right]$ is a plane wave with zero
average for all $\k \neq 0$ \cite{Fetter1972,Lewenstein1996}. These are crucial properties for the
low-energy excitations of the system to be well defined \cite{Gaul2011_bogoliubov_long}. 

As a second difference to the homogeneous case, 
these excitations now live on a deformed background. Both differences 
can be accounted for by defining a single effective 
potential $\V_{\k\k'} = \matrs{W}{Y}{Y}{W}_{\k\k'}$
that mediates scattering between the different components of
the Bogoliubov-Nambu 
pseudo spinor $\hat\Gamma_{\k}=
(\gh{\k},\ghd{-\k})^\text{t}$.  
From \eqref{eqBogoliubovHamiltonian}, we thus arrive at the inhomogeneous
Bogoliubov Hamiltonian 
\begin{equation}\label{eqInhomBgHamiltonian_Gamma}
\hat H = \frac{1}{2} \sum_{\k} \ep{k} \hat\Gamma_{\k}^\dagger\hat\Gamma_{\k} +
\frac{1}{2} \sum_{\k,\k'} 
\hat\Gamma^\dagger_{\k}
\V_{\k \k'}
\hat\Gamma_{\k'} 
\end{equation}
with the structure $\hat H = \hat H^{(0)} + \hat
H^{(V)}$. 
At this point, the only approximation made is the neglect of third and fourth
order terms in the fluctuations. 
In contrast, $ \hat H^{(V)}$ is still exact in the disorder
strength.

A perturbative, but fully analytical description is obtained by expanding 
$ \V  = \matrs{W}{Y}{Y}{W}= \V^{(1)} +  \V^{(2)} + \dots $ to
lowest orders in the bare disorder with the help of eqs.\ 
\eqref{eqSmoothingPsi1} and \eqref{eqSmoothingPsi2}. The small
parameter of this expansion is  $v=V/gn\ll 1 $. 
The first-order scattering amplitudes 
$W^{(1)}_{\k \k'} = w^{(1)}_{\k \k'} V_{\k-\k'}$ and $Y^{(1)}_{\k \k'}
= y^{(1)}_{\k \k'} V_{\k-\k'}$  
are proportional to $V_{\k-\k'}$, as required by conservation of momentum. All information
about interaction and condensate background is factorized into
amplitude envelopes 
\begin{align}
w^{(1)}_{\k \k'} &= 
\frac{a_ka_{k'}\xi^2 (1-\delta_{\k\k'})} {2+\xi^2(\k'-\k)^2}
\left[ k^2 + k'^2 -\k\cdot\k'
- \frac{ \k\cdot\k'}{a_k^{2} a_{k'}^{2}} \right],  \nonumber\\
y^{(1)}_{\k \k'} &= \frac{a_ka_{k'}\xi^2 (1-\delta_{\k\k'})} {2+\xi^2(\k'-\k)^2}
\left[ k^2 + k'^2 -\k\cdot\k'
+ \frac{ \k\cdot\k'}{a_k^{2} a_{k'}^{2}} \right]. 
 \label{w1y1}
\end{align}
Second-order scattering amplitudes are later only needed  for 
$\k=\k'$ because the disorder average  
restores translation invariance: 
$  W^{(2)}_{\k \k} = Y^{(2)}_{\k \k} 
 = \sum_{\p}   w^{(2)}_{\k \p}  V_{\k-\p} V_{\p-\k}$, 
with
\begin{equation} 
w^{(2)}_{\k\p}  = \frac{a_k^2\xi^2} {2gn}   \frac{p^2 + 3 (\k-\p)^2 +
  3  k^2}{[2+(\k-\p)^2\xi^2]^2} (1-\delta_{\k\p}).
\label{w2}
\end{equation}
This concludes our derivation of the inhomogeneous Bogoliubov
Hamiltonian. From here, one can derive numerous physical quantities
for any given potential $V(\r)$. 
For notational simplicity only, we assume in the following that $V(\r)$ describes 
disorder that is homogeneous and isotropic under the ensemble average, with
$\avg{V}=0$ and 
\begin{equation}\label{eqDisorderCorrelator}
\avg{V_\q V_{-\q'}} = L^{-d}\delta_{\q \q'} V^2 \sigma^d  C_d(q\sigma).
\end{equation}
The dimensionless function $C_d(q\sigma)$
characterizes the potential correlations persisting on the length
scale $\sigma$; the normalization is chosen such that in the
thermodynamic limit $\intdd{u} C_d(u) = 1$. 

%----------------------------------------
\section{Localization length}

The Hamiltonian \eqref{eqInhomBgHamiltonian_Gamma} is a random operator, varying with  
each realization of the quenched disorder potential. Therefore, the
Bogoliubov excitations in $d=1$ are expected to be localized
by the disorder \cite{Bilas2006,Lugan2007a}. 
And indeed, we can calculate
their localization length as $\lloc^{-1} =
\gamma_{2k}/2v_\text{g}$ from  the 
backscattering rate $\gamma_{2k}$ 
and group velocity
$\hbar v_\text{g}=\partial_k\ep{k}$.  To lowest order in the small parameter
$v=V/gn\ll 1 $, the backscattering rate derived by Fermi's Golden
Rule from the Hamiltonian
\eqref{eqInhomBgHamiltonian_Gamma} reads 
\begin{equation} 
\gamma_{2k} = 2\pi \rho(\ep{k})
\avg{W^{(1)\;2}_{k(-k)}},
\end{equation}
 where the density of states is 
$\rho(\ep{k})=[\pi\hbar v_\text{g}]^{-1}$. 
The resulting 
$\lloc^{-1} = \frac{1}{4} v^2 k^2\sigma  C_1(2 k\sigma)/(1+k^2\xi^2)^2$ 
agrees perfectly with \cite{Bilas2006,Lugan2007a}. 
Importantly, it is characteristic for sound waves that localization is
less pronounced at low energy. Indeed, for $k\xi\to 0$ at a fixed
correlation ratio $\zeta=\sigma/\xi$ of order unity, the 
localization rate per wave length vanishes like 
$[k\lloc]^{-1} \sim v^2 k\xi \to 0 $. Consequently, the low-energy
properties of the interacting quantum gas are not affected by
localization. The underlying reason is that the disorder  
is screened by interaction \cite{Lee1990}---contrary to the case
of noninteracting particles, where localization is stronger at lower
energy \cite{Kuhn2007a}. 

In higher dimensions, localization is even less pronounced, with the
localization length being exponentially large compared to the mean
free path, if not infinite. 
Low-energy excitations are free to propagate over long times and large
distances. The main effect of disorder then is to 
renormalize the excitation dispersion relation. 

%----------------------------------------
\section{Disorder-modified dispersion}

The disorder-modified quasiparticle dispersion
$\avg{\epsilon}_k=\ep{k}+\Delta\avg{\epsilon}_k$ can be determined by 
applying standard Nambu-Green perturbation theory \cite{Bruus2004}
to the relevant Hamiltonian \eqref{eqInhomBgHamiltonian_Gamma}. 
Thanks to its perturbative structure $\hat H = \hat H^{(0)} + \hat
H^{(V)}$, it is straightforward to calculate the self-energy of the
single-excitation Green function \cite{Gaul2011_bogoliubov_long}. The self-energy's
imaginary part provides the elastic scattering rate, which vanishes at
low energy just like the localization rate discussed above. From the
self-energy's real part, we deduce the shift in the dispersion 
\begin{align}\label{eqDeltaE}
 \frac{\Delta\avg{\epsilon}_k}{\epsilon_k} =  v^2\sigma^d \intdd{q} z_{\k \q} C_d(q\sigma),
\end{align}
with the kernel ($\PV$ denotes the principal value)
\begin{align}\label{eqKernelDeltaE}
  z_{\k \q}
 = \frac{g^2 n^2}{\epsilon_k} 
  \biggl[ \PV  \frac{[w^{(1)}_{\k (\k+\q)}]^2}{\epsilon_{k} - \epsilon_{\k+\q}}
 - \frac{[y^{(1)}_{\k (\k+\q)}]^2}{\epsilon_k + \epsilon_{\k+\q}} +
 w^{(2)}_{\k (\k+\q)} \biggr]. 
\end{align}
Together with expressions \eqref{w1y1} and \eqref{w2},
eq.~\eqref{eqDeltaE} allows calculating the dispersion of Bogoliubov
excitations in weak, but arbitrarily correlated disorder.  

In the hydrodynamic limit $\xi \to 0$, where the healing length is
shorter than both correlation length $\sigma$ and wavelength $2\pi/ k$,
eq.~\eqref{eqKernelDeltaE} simplifies considerably,
and eq.~\eqref{eqDeltaE} reproduces eq.~(28) of Ref.\cite{Gaul2009a}.
In this regime, 
the excitation energy is \emph{reduced}
in all dimensions and for any value of $k \sigma$.
For low energies $k\sigma \to 0$ and smooth potentials $\sigma\gg\xi$,
the sound-velocity shift $\Delta \avg{c}/c = -v^2/(2
d)$ is independent of the correlation details.

%------------------------------------------
\begin{figure}
\centerline{\includegraphics[width=0.95\linewidth]{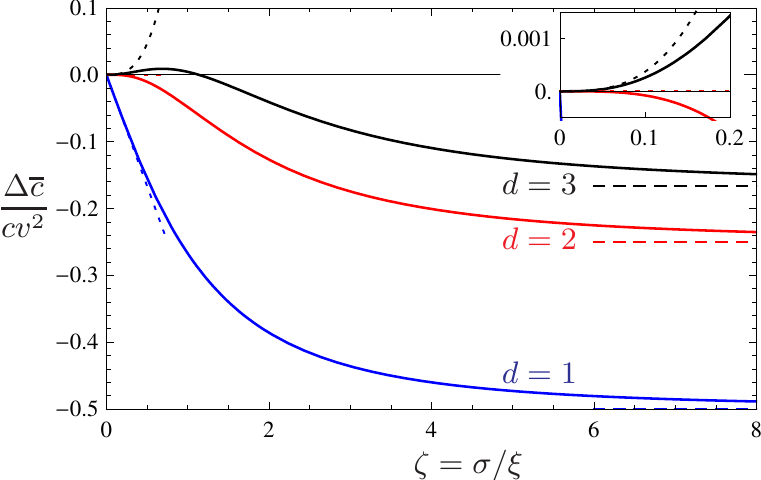}}
\caption{(Color online) Sound-velocity correction, computed
  from \eqref{eqDeltaE} and \eqref{eqEnvelopeK0}, due
  to Gaussian correlated disorder, eq.~\eqref{gausscorr.eq}, with variance $v^2:= V^2 /(gn)^2\ll1$ as function of 
correlation ratio  $\zeta=\sigma/\xi$. 
For most values, the sound velocity is reduced.
Dashed and dotted: universal limits for very
 smooth ($\zeta\gg 1$) and $\delta$-correlated disorder
($\zeta\ll1$) respectively, as collected in Tab.~\ref{tabDeltaCLimits}.
Inset: same data 
around the origin, showing the rapid departure from the leading-order 
estimate in $d=3$ \cite{Giorgini1994,Lopatin2002,Falco2007}.}
\label{figDeltaC_k0_gauss}
\end{figure}
%-----------------------------------------

We now change the point of view by taking the limit $k \to 0$, 
with an arbitrary correlation ratio
$\zeta = \sigma/\xi$.  
In particular, this allows us to reach the case of $\delta$-correlated disorder
where $\sigma \ll \xi, k^{-1}$. 
The kernel \eqref{eqKernelDeltaE} simplifies to
\begin{equation}\label{eqEnvelopeK0}
z_{0\q} = 2  \frac{q^2\xi^2 -
  (2+q^2\xi^2)\cos^2\beta}{(2+q^2\xi^2)^3},\quad \beta = \measuredangle(\k,\q).
\end{equation}

Fig.~\ref{figDeltaC_k0_gauss} shows the disorder correction
\eqref{eqDeltaE} to the speed of
sound, resulting from kernel
\eqref{eqEnvelopeK0} and a generic Gaussian
pair correlation 
\begin{equation} \label{gausscorr.eq}
C_d(q\sigma) = (2\pi)^{d/2} \exp\{-q^2\sigma^2/2\}.
\end{equation} 
 The plotted curves can
be expressed in closed form, but the details depend on the
specific correlator and are not of general
interest. 
In contrast, one finds universal %\old{limiting}
 behavior for very small or very large $\zeta$. 
The limit $\zeta \to \infty$ of a very smooth potential 
coincides, as it should, with the hydrodynamic limit 
$\Delta \avg{c}/c = -v^2/(2 d)$. 
In the opposite limit $\zeta \to 0$ of $\delta$-correlated disorder,
the 
correlator $C_d(0)$ can
be pulled out of the integral \eqref{eqDeltaE},
which becomes elementary. 
The correction then scales as 
$\zeta^d$, 
as shown in Fig.~\ref{figDeltaC_k0_gauss}, with prefactors that are
collected in Tab.~\ref{tabDeltaCLimits}.  
\footnote{In $d=2$, the universal coefficient in front of $\zeta^2$ is
  zero. For the Gaussian correlation \eqref{gausscorr.eq}, the 
first finite term is $\frac{1}{4} \zeta^4 (2 \ln \zeta + 1 + \gamma)$
with  Euler's constant  $\gamma$\label{Note2D}.}
Notably, we corroborate the  
known result for $\delta$-correlated disorder in $d=3$ 
\cite{Giorgini1994,Lopatin2002,Falco2007},  
the only case with a positive correction.
But our theory reveals that this estimate is
of limited use because already a small correlation makes a
large difference, as shown by the inset of
Fig.~\ref{figDeltaC_k0_gauss}.

%---------------------------------------------
\begin{table}
\begin{center}
 \begin{tabular}{c|ccc}
  $\Delta \avg{c}/cv^2
\vphantom{\displaystyle{\sum}}$     & $d=1$              & $d=2$                  & $d=3$\\
\hline
$\zeta\gg1
\vphantom{\displaystyle{\sum}}$     & $-\frac{1}{2}$  & $-\frac{1}{4} $ & $-\frac{1}{6}$\\[5pt]
$\zeta \ll 1
\vphantom{\displaystyle{\sum}}$     & $-\frac{3C _1(0) \zeta}{16\sqrt{2}} $&   $0$    & $+\frac{5C_3(0)
  \zeta^3}{48\sqrt{2}\pi} $ 
\end{tabular}
\caption{Universal limits of the speed-of-sound correction, computed
  from \eqref{eqDeltaE} and \eqref{eqEnvelopeK0}, for very
  smooth disorder ($\zeta\gg 1$) and $\delta$-correlated disorder
  ($\zeta\ll1$). See also footnote \ref{Note2D}.}
\label{tabDeltaCLimits}
\end{center}
\end{table}

%----------------------------------------
\section{Condensate depletion}

Finally, we investigate the condensate depletion properly speaking,
namely the density of particles out of the mean-field
condensate, 
$\delta n :=
L^{-d}\intddr
\xpct{\delta\hat\Psi(\r)^\dagger\delta\hat\Psi(\r)}$. 
Inserting \eqref{eqBasedOnHydro} and rearranging terms, we find that
the depletion density can be written   
\begin{align}\label{depletion1.eq}
&\delta n  = \frac{1}{4nL^d} \sum_{\k,\k'} 
 \left\{ \left[ a_k a_{k'} \check n_{\k'-\k} + \frac{ n_{\k'-\k}}{a_ka_{k'}}\right] 
\xpct{\gh{\k}\ghd{\k'} +\ghd{\k'}\gh{\k}}  
\right. \nonumber
\\ 
& 
+ \left. \left[a_k a_{k'}\check n_{\k'-\k} -  \frac{n_{\k'-\k}}{a_ka_{k'}}\right] 
\xpct{\gh{\k}\gh{-\k'} +\ghd{\k'}\ghd{-\k}} - 2 
\delta_{\k\k'}  
\right\}. 
\end{align}
In principle, this expression is correct to all orders in $V$.  
One only requires the Fourier components  
$n_{\k}:= [\Phi(\r)^2]_\k  = L^{-d} \sum_\q \Phi_{\k-\q}\Phi_\q$
of the deformed mean-field condensate density, as well as the
 Fourier components of its inverse, 
$\check n_{\k}:=n^2[\Phi(\r)^{-2}]_{\k}$. 
Up to order $V^2$, they 
follow from the smoothing-theory results \eqref{eqSmoothingPsi1} and
\eqref{eqSmoothingPsi2}. 
One also has to compute the 
Bogoliubov expectation values to the desired order. 
With the inhomogeneous Bogoliubov Hamiltonian
\eqref{eqInhomBgHamiltonian_Gamma} at hand,  
this is again a standard task in perturbation
theory using the Nambu formalism
\cite{Bruus2004}.   

Let us first check the homogeneous case $V=0$. 
Then, \eqref{depletion1.eq} shrinks to 
$\delta n^{(0)} = \frac{1}{4} \intdd{k} 
\left[a_k -  a_k^{-1}  
\right]^2$, a well-known result \cite{Pitaevskii2003}. 
In $d=3$, this evaluates to a depletion density 
$\delta n^{(0)} = [6\sqrt{2}\pi^2 \xi^3]^{-1} $
or equivalently to the relative depletion 
$\delta n^{(0)}/n = 8 (n a_s^3)^{1/2} / 3\sqrt{\pi} $. 
In $d=2$, one finds 
$\delta n^{(0)} = [8\pi \xi^2]^{-1}$.  
The $d=1$ integral is infrared divergent, consistent with the
fact that zero-point fluctuations prevent homogeneous 1D BECs.  
Cutting off the integral at
some value $\aIR=\xi k_\text{IR}\ll1$, with $k_\text{IR}$ of the order
of the inverse system size, one finds
$\delta n^{(0)} = (2\ln2-2-\ln\aIR)/(2\sqrt{2}\pi \xi) $, up to order
$\aIR$.

\begin{figure}
\centerline{\includegraphics[width=\linewidth]{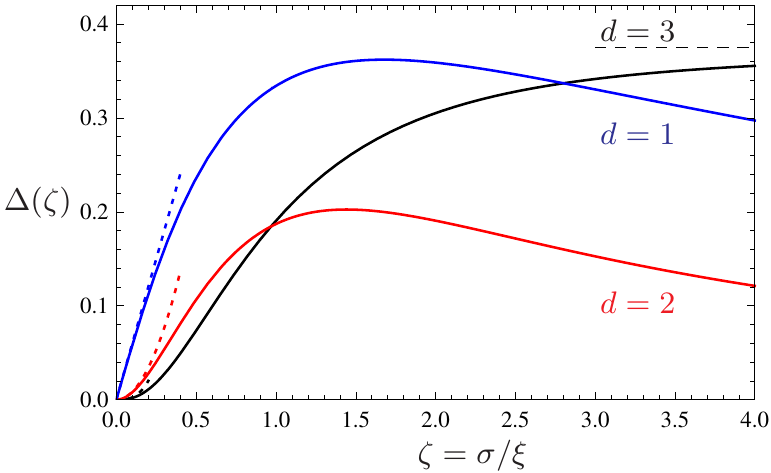}}
\caption{
Disorder-induced condensate depletion~\eqref{Deltazeta}, relative to the clean value and
in units of disorder strength $ v^2$, as function of
the correlation ratio $\zeta=\sigma/\xi$ for the Gaussian
correlation \eqref{gausscorr.eq}.
Dashed and dotted: universal limits as collected in
Tab.~\ref{tabDeltanlimits}.} 
\label{deln_eta_0gaussd123.fig} 
\end{figure}

Now we evaluate the disorder-induced depletion by expanding
all contributions to \eqref{depletion1.eq} to second order in $V$. 
Upon taking the ensemble average, 
terms of order $V$ average to zero, and 
$\avg{\delta n}=\delta n^{(0)}+\avdn{2}+O(v^3)$. 
Each of
the second-order terms is proportional to the correlator
\eqref{eqDisorderCorrelator}, and we can collect all contributions into a
single kernel: 
\be\label{Deltazeta}
 \avdn{2}=  v^2  \delta n^{(0)}  
\int \frac{\rmd^dq}{(2\pi)^d} \sigma^d C_d(q\sigma) G_d(q\xi). 
\ee
Details of this derivation will be given elsewhere
\cite{Muller2011}. Here we note that the relative depletion in units
of $v^2$,  $\Delta(\zeta) = \avdn{2}/v^2\delta n^{(0)}$  is only
function of $\zeta=\sigma/\xi$.
\footnote{Only in $d=1$, it depends also weakly on the cutoff $\aIR$ that 
regularizes already the clean depletion; plots in this paper are done
with $\alpha=0.01$. We stress that our calculations require
  \emph{no additional ad-hoc cutoffs},
neither infrared (since the excitations are orthogonal to the
vacuum) nor ultraviolet (since potential correlations
are included)\label{Note1d}.} 
This quantity is 
plotted in  Fig.~\ref{deln_eta_0gaussd123.fig}. 
As for the sound velocity correction, details depend on
the specific correlation.  
In the
$\delta$-correlated limit $\zeta\to0$,
Fig.~\ref{deln_eta_0gaussd123.fig} shows the generic
scaling $\Delta(\zeta) = \beta_d \zeta^d C_d(0)$ with 
numerical coefficients $\beta_1\approx 0.235$ (weakly dependent on the
cutoff $\alpha$), $\beta_2\approx 0.135$,
$\beta_3\approx 0.160$. In the limit  $\zeta\to\infty$ of a
very smooth potential, we find $\Delta \to G_d(0)$ with  $G_3(0)=3/8$,
$G_2(0)=0$, and $G_1(0)=-1/8$. In all cases but the last, the
condensate depletion due to disorder is positive, as expected. As
shown in Fig.~\ref{deln_eta_0gaussd123.fig}, also in $d=1$ the
depletion is positive for most values of $\zeta$. The curve only
crosses over to negative values for such a large value $\zeta=\sigma/\xi$
depending on the cutoff $\alpha$, that the correlation length $\sigma$
has to be comparable to the system size, which is not the regime of
present interest.

%------------------------
%---------------------------------------------
\begin{table}
\begin{center}
 \begin{tabular}{c|ccc}
  $\Delta(\zeta)
\vphantom{\displaystyle{\sum}}$   \,&\, $d=1$               & $d=2$                  & $d=3$\\
\hline
$\zeta\gg1
\vphantom{\displaystyle{\sum}}$   \,&\, $-\frac{1}{8}$      & $ 0  $ & $\frac{3}{8}$\\
$\zeta \ll 1
\vphantom{\displaystyle{\sum}}$   \,&\, $\beta_1 C_1(0) \zeta $ & $ \beta_2 C_2(0)\zeta^2 $    &
$\beta_3 C_3(0)\zeta^3 $ 
\end{tabular}
\caption{Universal limits of the disorder-induced condensate
  depletion, relative to the clean value and in units of $v^2$ as plotted in
  Fig.~\ref{deln_eta_0gaussd123.fig}, for very
  smooth disorder ($\zeta\gg 1$) and $\delta$-correlated disorder
  ($\zeta\ll1$). For the latter case, the numerical coefficients are 
$\beta_1\approx 0.235$ (for $\alpha=0.01$), $\beta_2\approx 0.135$,
$\beta_3\approx 0.160$.}
\label{tabDeltanlimits}
\end{center}
\end{table}
%----------------------------------------
 
In all cases, the combined depletion due to
interaction and disorder reads 
$\avg{\delta n}=\delta n^{(0)}[1 + v^2 \Delta(\zeta)]$, with
$\Delta(\zeta)$ at most of the order of unity.  
Clearly, the fractional depletion induced by the disorder,  
$\avdn{2}/n = (\delta n^{(0)}/n) v^2 \Delta(\zeta)$ is a factor 
$\delta n^{(0)}/ n\ll1$ smaller than the mean-field condensate
deformation, which is of order $v^2$. In hindsight, this result is
rather plausible: the primary effect of the external disorder potential is
merely to deform the condensate. The depletion of the
condensate itself is a secondary scattering effect, mediated by the
weak repulsive boson interaction, and therefore considerably weaker. 

In conclusion, we report substantial progress in the analytical description of
interacting, condensed bosons in correlated disorder of any dimensionality. We derive the
fundamental Bogoliubov Hamiltonian for excitations. This determines a
wealth of (thermo-)dynamic quantities, out of which we calculate the sound velocity in
all dimensions. Moreover, we calculate the disorder-induced quantum
depletion, 
which proves to be much smaller 
than the previously known mean-field condensate
deformation. We conclude that our 
theory should fare very well in
describing the excitations of disordered interacting bosons, especially in dilute
cold gases, where the study of well-controlled disorder in earnest has 
just begun \cite{Sanchez-Palencia2010,Modugno2010,Dries2010}.

\begin{acknowledgments}
This work is supported by the National Research Foundation \& Ministry of
Education, Singapore, and the Spanish MEC (Project MOSAICO). Financial support by Deutsche
Forschungsgemeinschaft is acknowledged for the time when
both authors were affiliated with Universit\"at Bayreuth,
Germany.
We are grateful for helpful discussions with
T.~Giamarchi, 
V.~Gurarie, 
P.~Lugan, 
A.~Pelster,  
L.~Sanchez-Palencia,
and E.~Zaremba. 
\end{acknowledgments}

\bibliography{../literatureBEC}

\end{document}